\documentclass[10pt]{article}
\usepackage{epsfig}

\newif\ifdimspec
\def\figsize#1{\dimspecfalse \checkdim#1\end
\ifdimspec
  \def\figureWidth{#1}%
\else
  \def\figureWidth{#1 in}\fi}
\def\checkdim#1{\ifx#1\end \let\next=\relax
  \else \ifcat#1a \dimspectrue \fi \let\next=\checkdim\fi \next}
\newcommand{\lblcaption  }[2]{\caption{#2\label {#1}}}
\newcommand{\Figure }[2]{\begin{figure} 
  \begin{center}
  \mbox{\epsfig{file=#1.eps}}
  \end{center}
  \lblcaption{#1}{#2}
  \end{figure}}
\newcommand{\Figuresized }[3]{\begin{figure} 
  \begin{center}
  \mbox{\figsize{#3}\epsfig{file=#1.eps,width=\figureWidth}}
  \end{center}
  \lblcaption{#1}{#2}
  \end{figure}}
\newcommand{\twoAngFiguresEPS}[6]
{
\begin{figure}
\begin{center}
\figsize{#4}
\begin{minipage}[t]{3.2in}
\begin{center}
\epsfig{file=#1.eps,width=\figureWidth,angle=#5}
\end{center}
\end{minipage}
\begin{minipage}[t]{3.2in}
\begin{center}
\epsfig{file=#2.eps,width=\figureWidth,angle=#6}
\end{minipage}
\end{center}
\lblcaption{#1}{#3}
\end{figure}
}
\DeclareMathAlphabet{\mathbfit}{OT1}{cmr}{bx}{it}

\makeatletter

\def\singlespace{\vskip\parskip
\vskip\baselineskip
\def\baselinestretch{1}
\ifx\undefined\selectfont 
\ifx\@currsize\normalsize\@normalsize\else\@currsize\fi
\else
\@newbaseline\fi 
\vskip-\parskip
\vskip-\baselineskip}

\let\fancy@def\gdef
\def\lhead{\@ifnextchar[{\@xlhead}{\@ylhead}}
\def\@xlhead[#1]#2{\fancy@def\@elhead{#1}\fancy@def\@olhead{#2}}
\def\@ylhead#1{\fancy@def\@elhead{#1}\fancy@def\@olhead{#1}}
\def\chead{\@ifnextchar[{\@xchead}{\@ychead}}
\def\@xchead[#1]#2{\fancy@def\@echead{#1}\fancy@def\@ochead{#2}}
\def\@ychead#1{\fancy@def\@echead{#1}\fancy@def\@ochead{#1}}
\def\rhead{\@ifnextchar[{\@xrhead}{\@yrhead}}
\def\@xrhead[#1]#2{\fancy@def\@erhead{#1}\fancy@def\@orhead{#2}}
\def\@yrhead#1{\fancy@def\@erhead{#1}\fancy@def\@orhead{#1}}
\def\lfoot{\@ifnextchar[{\@xlfoot}{\@ylfoot}}
\def\@xlfoot[#1]#2{\fancy@def\@elfoot{#1}\fancy@def\@olfoot{#2}}
\def\@ylfoot#1{\fancy@def\@elfoot{#1}\fancy@def\@olfoot{#1}}
\def\cfoot{\@ifnextchar[{\@xcfoot}{\@ycfoot}}
\def\@xcfoot[#1]#2{\fancy@def\@ecfoot{#1}\fancy@def\@ocfoot{#2}}
\def\@ycfoot#1{\fancy@def\@ecfoot{#1}\fancy@def\@ocfoot{#1}}
\def\rfoot{\@ifnextchar[{\@xrfoot}{\@yrfoot}}
\def\@xrfoot[#1]#2{\fancy@def\@erfoot{#1}\fancy@def\@orfoot{#2}}
\def\@yrfoot#1{\fancy@def\@erfoot{#1}\fancy@def\@orfoot{#1}}
\newdimen\headrulewidth
\newdimen\footrulewidth
\newdimen\plainheadrulewidth
\newdimen\plainfootrulewidth
\newdimen\headwidth
\newdimen\footruleskip
\newif\if@fancyplain \@fancyplainfalse
\def\fancyplain#1#2{\if@fancyplain#1\else#2\fi}
\def\fancy@reset{\restorecr
 \def\baselinestretch{1}%
 \def\nouppercase##1{{\let\uppercase\relax\let\MakeUppercase\relax##1}}%
 \ifx\undefined\@newbaseline
   \ifx\@normalsize\undefined \normalsize 
   \else \@normalsize \fi
 \else
  \@newbaseline%
 \fi}
\footrulewidth\z@
\plainheadrulewidth\z@
\plainfootrulewidth\z@
\def\@fancyhead#1#2#3#4#5{#1\hbox to\headwidth{\fancy@reset\vbox{\hbox
{\rlap{\parbox[b]{\headwidth}{\raggedright#2\strut}}\hfill
\parbox[b]{\headwidth}{\centering#3\strut}\hfill
\llap{\parbox[b]{\headwidth}{\raggedleft#4\strut}}}\headrule}}#5}
\def\@fancyfoot#1#2#3#4#5{#1\hbox to\headwidth{\fancy@reset\vbox{\footrule
\hbox{\rlap{\parbox[t]{\headwidth}{\raggedright#2\strut}}\hfill
\parbox[t]{\headwidth}{\centering#3\strut}\hfill
\llap{\parbox[t]{\headwidth}{\raggedleft#4\strut}}}}}#5}
\def\headrule{{\if@fancyplain\let\headrulewidth\plainheadrulewidth\fi
\hrule\@height\headrulewidth\@width\headwidth \vskip-\headrulewidth}}
\def\footrule{{\if@fancyplain\let\footrulewidth\plainfootrulewidth\fi
\vskip-\footruleskip\vskip-\footrulewidth
\hrule\@width\headwidth\@height\footrulewidth\vskip\footruleskip}}
\def\ps@fancy{%
\@ifundefined{@chapapp}{\let\@chapapp\chaptername}{}
\@ifundefined{MakeUppercase}{\def\MakeUppercase{\uppercase}}{}
\@ifundefined{chapter}{\def\sectionmark##1{\markboth
{\MakeUppercase{\ifnum \c@secnumdepth>\z@
 \thesection\hskip 1em\relax \fi ##1}}{}}%
\def\subsectionmark##1{\markright {\ifnum \c@secnumdepth >\@ne
 \thesubsection\hskip 1em\relax \fi ##1}}}%
{\def\chaptermark##1{\markboth {\MakeUppercase{\ifnum \c@secnumdepth>\m@ne
 \@chapapp\ \thechapter. \ \fi ##1}}{}}
\def\sectionmark##1{\markright{\MakeUppercase{\ifnum \c@secnumdepth >\z@
 \thesection. \ \fi ##1}}}}%
\ps@@fancy
\gdef\ps@fancy{\@fancyplainfalse\ps@@fancy}%
\ifdim\headwidth<0sp
    \advance\headwidth123456789sp\advance\headwidth\textwidth\fi}
\def\ps@fancyplain{\ps@fancy \let\ps@plain\ps@plain@fancy}
\def\ps@plain@fancy{\@fancyplaintrue\ps@@fancy}
\def\ps@@fancy{%
\ps@empty 
\def\@mkboth{\protect\markboth}%
\def\@oddhead{\@fancyhead\@lodd\@olhead\@ochead\@orhead\@rodd}%
\def\@oddfoot{\@fancyfoot\@lodd\@olfoot\@ocfoot\@orfoot\@rodd}%
\def\@evenhead{\@fancyhead\@rodd\@elhead\@echead\@erhead\@lodd}%
\def\@evenfoot{\@fancyfoot\@rodd\@elfoot\@ecfoot\@erfoot\@lodd}%
}
\def\@lodd{\if@reversemargin\hss\else\relax\fi}
\def\@rodd{\if@reversemargin\relax\else\hss\fi}
\let\latex@makecol\@makecol
\def\@makecol{\let\topfloat\@toplist\let\botfloat\@botlist\latex@makecol}
\def\iftopfloat#1#2{\ifx\topfloat\empty #2\else #1\fi}
\def\ifbotfloat#1#2{\ifx\botfloat\empty #2\else #1\fi}
\def\iffloatpage#1#2{\if@fcolmade #1\else #2\fi}
\newcommand{\fancypagestyle}[2]{%
  \@namedef{ps@#1}{\let\fancy@def\def#2\relax\ps@fancy}}
\makeatother
\begin{document}
\baselineskip=15pt

\vbox to 1cm{\vfill}

\lhead[]{\fontsize{9pt}{15pt}\selectfont\rm 
F.~De Martini, G.~Denardo, and Y.~Shih \rm (eds.), \it Quantum 
Interferometry\/\rm,\\ VCH Publishing Division I, New York (1996), 
pp.~193--204.}
\lfoot{}
\cfoot{}
\rfoot{}
\setlength{\headrulewidth}{0pt}
\setlength{\footrulewidth}{0pt}

\begin{center}
{\fontsize{14pt}{14pt}\selectfont\bf Event--Ready Entanglement Preparation} 
\end{center}

\begin{center}
{\it Mladen Pavi\v ci\'c}
\end{center}

\begin{center}
{Max--Planck--AG Nichtklassische Strahlung,  
Humboldt Univ., Berlin, Germany\\
and University of Zagreb, Croatia; \\ E--mail: mpavicic@faust.irb.hr; 
Web page: http://m3k.grad.hr}
\end{center}

\renewcommand{\abstractname}{}
\begin{abstract}
\bf Abstract. \rm All Bell experiments carried out so far have had 
ten or more times fewer coincidence counts than singles counts and 
this, in effect, means a detection efficiency under 10\%. Therefore, 
all these experiments relied only on coincidence counts and herewith 
on additional assumptions. Recently, however, Santos devised 
hidden variable models which do not obey the assumptions and thus 
made the experiments inconclusive. This, as well as recent 
improvements in detectors efficiencies, prompted an increasing 
interest in the \it loophole--free\/ \rm Bell experiments which 
do not rely on additional assumptions and which originally stem 
from the idea of the \it event--ready\/ \rm detectors (introduced by 
J.S.~Bell) which would preselect Bell pairs ready for detection.  
Till recently it was assumed that such detectors would  
distort the pairs. Here we devise those that would not do so and propose 
an experiment which can realistically improve the detection efficiency 
and visibility up to over 80\%. The set--up uses two nonlinear crystals 
of type--II both of which simultaneously downconvert a singlet--like 
pair. We combine one photon from the first singlet with one 
from the second singlet at a beam splitter and consider their 
coincidence detections. Detectors determine optimally narrow 
solid angles for the downconverted photons. However, for their 
two companions (from each singlet) we use 
five times wider solid angles or even drop pinholes altogether and 
resort to frequency filters. So, we are able to 
realistically collect close to 100\%\ of them. The latter 
pairs---preselected by coincidence detection at the beam 
splitter---appear entangled in (non)maximal singlet--like states, 
i.e., detectors at the beam splitter act as event--ready detectors 
for such Bell pairs.

\end{abstract}

\bigskip
\bigskip

\section{\fontsize{12pt}{12pt}\selectfont\bf Introduction}
\label{sec:Intro}

Although many convincing EPR (Einstein--Podolsky--Rosen) 
experiments violating the local hidden variable models and 
various forms of Bell inequalities  were performed in the 
past thirty years, an experiment involving no 
supplementary assumptions---usually called a \it 
loophole--free\/ \rm experiment---is still waiting to be 
carried out. Until recently loophole--free experiments were 
not considered because they require very high detection 
efficiency~[4] and all experiments carried out till 
now have had an efficiency under 10\%\ $\!$[10,15]. 
On the other hand, the most important supplementary 
assumption, the \it no enhancement assumption\/ \rm and 
the corresponding postselection method were  
considered to be very plausible. Then Santos 
devised~[22--25] local hidden--variable 
models which violate not only the low detection loophole but also the 
\it no enhancement assumption\/ \rm as well as post--selection loophole, 
and these models, as well as considerable improvements in techniques, 
in particular, detector efficiencies, resulted in an interest into 
\it loophole--free\/ \rm experiments. In the past two 
years several sophisticated proposals appeared 
which rely on the recent improvement in the detection technology and 
meticulous elaborations of all experimental details.~[6,11,12,14,18] 
The first three use maximal 
superpostions and require detection efficiency of at least 
83\%\ $\!$[7] and the other two use nonmaximal 
superpositions relying on recent results~[5,19,20] which require 
only 67\%\ detection efficiency for them. 
All proposals are very demanding and at the same time all but the last 
proposal invoke a postselection which is also a supplementary 
assumption.~[25] In this paper we analyze several  
supplementary assumptions and propose a feasible method of doing a 
loophole--free Bell experiment which requires only 67\%\ detection 
efficiency, can work with a realistic visibility, and uses 
a preselection method for preparing non--maximally entangled photon 
pairs. The preselection method is particularly attractive for its  
ability to employ solid angles of signal and idler photons
(in a downconversion process in a nonlinear crystal) which differ up 
to five times from each other. This enables a tremendous increase 
in detection efficiency---from 10\%\ to over 80\%---as elaborated below.  

\lhead[]{\fontsize{9pt}{15pt}\selectfont\bf
194\hfill Mladen Pavi\v ci\'c}
\lfoot{}
\cfoot{}
\rfoot{}
\setlength{\headrulewidth}{0pt}
\setlength{\footrulewidth}{0pt}

\bigskip

\section{\fontsize{12pt}{12pt}\selectfont\bf Bell inequalities and 
their supplementary assumptions}
\label{sec:Bell}

As we mentioned in the introduction the recent revival of the Bell 
issue has been partly triggered by new types of local hidden 
variables devised by Santos~[22,25] which made all 
experiments carried out so far inconclusive. However, 
from the very first Bell experiments it was clear that one day  
a conclusive loophole--free experiment must be carried 
out.~[3,4] At the time, such experiments were far from being 
feasible and as a consequence all experiments so far relied on 
coincidental detections and on an assumption that \it a subset of 
a total set of events would give the same statistics as the set 
itself\/\rm. In other words no real experiment so far dealt with 
proper probabilities, i.e., with \it ratios of detected events 
to copies of the physical system initially prepared\/\rm.~[12] 
Let us see this point in some more details, first, for the 
Clauser--Horne~[3] form of the Bell inequality, and then for 
Hardy's equality.~[9]

We consider a composite system containing two subsytems in a 
(non)maximal superposition. When a property is being measured on 
subsystem $i$ by detector D$_i$, which has got an adjustable 
parameter $a_i$ corresponding to the property, the probability 
of an independent firing of one of the two detectors is 
$p(a_i)=N(a_i)/N$ ($i=1,2$) and of simultaneous triggering of both 
detectors is $p(a_1,a_2)=N(a_1,a_2)/N$, where $N(a_i)$ is the 
number of counts at D$_i$, $N(a_1,a_2)$ is the number of 
coincident counts, and $N$ is the total number of the systems 
the source emits. Let a classical hidden state $\lambda$ determine 
the individual counts and the 
probabilities of individual subsystems triggering the detectors: 
$p(\lambda,a_i)$ and $p(\lambda,a_1,a_2)$. 
These probabilities are connected with the above introduced \it 
long run\/ \rm probabilities by means of: $p(a_i)=\int_\Gamma
\rho(\lambda)p(\lambda,a_i)d\lambda$ ($i=1,2$) and 
$p(a_1,a_2)=\int_\Gamma\rho(\lambda)p(\lambda,a_1,a_2)d\lambda$, 
where $\Gamma$ is the space of states $\lambda$ and $\rho(\lambda)$ 
is the normalized probability density over states $\lambda$. 
The locality condition---which assumes that the probability of 
one of the detectors being triggered does not depend on whether 
the other one has been triggered or not---can be formalized as 
$p(\lambda,a_1,a_2)=p(\lambda,a_1)p(\lambda,a_2)$.
Clauser--Horne's form of the Bell inequality reads:
\begin{eqnarray}
-A_1\>A_2\leq p(\lambda,a_1)p(\lambda,a_2)-
p(\lambda,a_1)&\!\!\!\!\!\!p(\lambda,a_2')+
p(\lambda,a_1')p(\lambda,a_2')+p(\lambda,a_1')p(\lambda,
a_2)\nonumber\\
&-\,A_2\,p(\lambda,a_1')-A_1\,p(\lambda
,a_2)\leq0\,,\!\!\!\!\!\label{eq:clauser} 
\end{eqnarray} 
where $0\le p(\lambda,a_i)\le A_i$. 

\lhead[]{\fontsize{9pt}{15pt}\selectfont\bf
Event--Ready Entanglement Preparation\hfill 195}
\chead{}
\rhead[]{}
\lfoot{}
\cfoot{}
\rfoot{}
\setlength{\headrulewidth}{0pt}
\setlength{\footrulewidth}{0pt}

The experiments carried out so far invoked the  
\it no--enhancement\/ \rm assumption $A_i=p(\lambda,\infty)$   
(where $\infty$ means that a filter for a property 
corresponding to parameters $a_i$ is switched off), wherewith 
Eq.~(\ref{eq:clauser}) after multiplication by 
$\rho(\lambda)$ and integration over $\lambda$ yields
\begin{equation}
-1\leq {p(a_1,a_2)\over p(\infty,\infty)}-
{p(a_1,a_2')\over p(\infty,\infty)}+{p(a_1',a_2')\over p(\infty,\infty)}+
{p(a_1',a_2)\over p(\infty,\infty)}-{p(a_1',\infty)\over p(\infty,\infty)}-
{p(\infty,a_2)\over p(\infty,\infty)}\leq0\label{eq:enh} 
\end{equation} 
Thus---because of the low detection efficiency---all 
the experiments performed till now measured nothing but the above 
ratios. Then Santos devised~[22,23,25] hidden variables
based on $p(\lambda,a_i)>p(\lambda,\infty)$ and left us only with 
the \it loophole--free\/ \rm option $A_1=A_2=1$ wherewith 
Eq.~(\ref{eq:clauser}) yields
\begin{eqnarray}
-1\leq p(a_1,a_2)-p(a_1,a_2')+p(a_1',a_2')+
p(a_1',a_2)-p(a_1')-p(a_2)\leq0\,.\label{eq:loop} 
\end{eqnarray} 

The above cited loophole--free proposals used the right 
inequality which requires 83\%\ detection efficiency for maximal 
superpositions and 67\%\ detection efficiency for nonmaximal 
ones. The left inequality always requires 83\%\ detection 
efficiency but it makes clear that if we want a \it 
loophole--free\/ \rm experiment we must always either register 
or preselect practically all the systems the source emits 
in order to obtain \it proper\/ \rm probabilities, i.e., \it ratios 
of detected events to the number of emitted systems\/\rm. 
An excellent test which immediately shows whether a particular 
experiment can be loophole--free is to see whether we can 
obtain $p(a_1)\approx p(a_1,\pm)\approx p(a_1,\infty)$,      
where `$\pm$' means that a two--channel filter (corresponding to 
a property $a$ and property non--$a$), e.g., a birefringent prism,  
is used; `$\infty$' means that the filter has been taken out 
altogether. Unfortunately all experiments carried out so far 
have $p(a_1)>10\,p(a_1,\infty)$. This applies to other approaches 
as well. E.g., Ardehali's additional assumptions~[1,2] 
are weaker than the \it no enhancement assumption\/ \rm but that 
does not help us in obtaining the proper probabilities. The 
latter is also true for the Hardy's equality experiment recently 
carried  out by Torgerson, Branning, Monken, and Mandel~[27] 
although they misleadingly claim that their ``method does not depend 
on the use of detectors with high or even known quantum 
efficiencies.''~[26] Let us look at the experiment in some detail. 

\lhead[]{\fontsize{9pt}{15pt}\selectfont\bf
196\hfill Mladen Pavi\v ci\'c}
\lfoot{}
\cfoot{}
\rfoot{}
\setlength{\headrulewidth}{0pt}
\setlength{\footrulewidth}{0pt}

Torgerson, Branning, Monken, and Mandel argue, in effect, as follows. 
In a two--photon polarization coincidence experiment at an asymmetric 
beam splitter one can---assuming 100\%\ efficiency---pick up the 
orientation angles of the polarizers so as to have 
$P(\theta_1,\theta_2')/P(\theta_1)=1$ and 
$P(\theta_1',\theta_2)/P(\theta_2)=1$, i.e., polarization 
$\theta_1$ must occur together with $\theta_2'$ and $\theta_2$ with 
$\theta_1'$. Classically, if $\theta_1$ and $\theta_2$ sometimes occur 
together, then $\theta_2'$ and $\theta_1'$ should also sometimes occur 
together. In a quantum measurement though, for a particular 
reflectivity of the beam splitter one can ideally obtain 
$P(\theta_1,\theta_2)>0$ together with $P(\theta_1',\theta_2')=0$ 
which is a contradiction for a classical reasoning. When detection
efficiency is far bellow 100\%\ one can assume that only coincidence 
data are relevant and substitute $P(\theta_1,\theta_2^\perp)+
P(\theta_1,\theta_2)$ for $P(\theta_1)$. If we define 
$P(\theta_1,\theta_2)=N(\theta_1,\theta_2)/
[N(\theta_1,\theta_2')+N(\theta_1,\theta_2'^\perp)+
N(\theta_1'^\perp,\theta_2')+N(\theta_1^\perp,\theta_2'^\perp)]$, where 
$N$'s are two--photon coincidence detections, for the considered 
experiment we arrive at 98\%\ efficiency. But,  
in doing so, we disregard first, that $2R(1-R)$ percent 
(44\%\ for the chosen $R$) of photons emerge from the same sides of  
the beam splitter, and secondly, that for the chosen source (LiIO$_3$ 
type--I downconverter) one has 
$P(\theta_1)>20[P(\theta_1,\theta_2^\perp)+
P(\theta_1,\theta_2)]$.~[16] Thus, we end up 
not with $P(\theta_1,\theta_2')/P(\theta_1)\approx 1$ but with 
$P(\theta_1,\theta_2')/P(\theta_1)=0.02$. In other words, the 
experiment is not a candidate for the loophole--free type of Bell 
experiments although it is one of the most convincing coincidence 
counts experiments carried out so far. 

\bigskip

\section{\fontsize{12pt}{12pt}\selectfont\bf Experiment}
\label{sec:experiment}

A schematic representation of the experiment is shown in
Fig.~1. Two independent type--II crystals (BBO)  
act as two independent sources of two independent singlet pairs. 
Two photons from each pair interfere at an asymmetrical beam splitter, 
BS and whenever they emerge from its opposite sides, pass through 
polarizers P1' and P2', and fire the detectors D1' and D2', they open 
the gate (activate the Pockels cells) which preselects the other two 
photons into a nonmaximal singlet state. We achieve the high efficiency 
(over 80\%) by choosing optimally narrow solid angles determined by 
the openings of D1' and D2' and five times wider solid angles determined
by D1 and D2. [Type--II crystal, as a source of only one singlet 
pair~[8,15], suffers from low efficiency (at most 
10\%\ $\!$[15]) due to necessarily symmetric detector solid angles.] 

\Figure{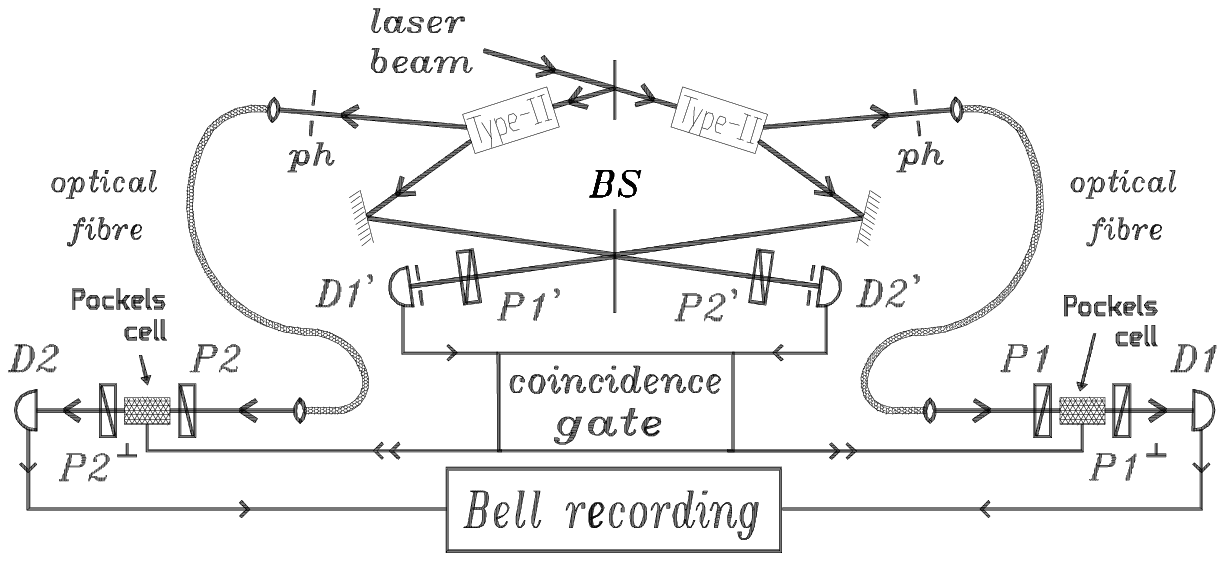}{Lay--out of the proposed experiment. 
As the event--ready preselector serves a beam splitter with detectors 
D1' and  D2' and their counters which open the gate (Pockells cells) 
for Bell pairs.}

\lhead[]{\fontsize{9pt}{15pt}\selectfont\bf
Event--Ready Entanglement Preparation\hfill 197 }
\chead{}
\rhead[]{}
\lfoot{}
\cfoot{}
\rfoot{}
\setlength{\headrulewidth}{0pt}
\setlength{\footrulewidth}{0pt}

\medskip 

 An ultrashort laser beam (a subpicosecond one) of 
frequency $\omega_0$ simultaneously 
(split by a beam splitter) pumps up two nonlinear crystals of 
type--II producing in each of them intersecting cones of 
mutually perpendicularly polarized signal and idler photons of 
frequencies $\omega_0/2$ as shown in Fig.~2. The idler 
and signal photon pairs coming out from the crystals do not have 
definite phases and therefore cannot exhibit second order 
interference. However they do appear entangled along the cone 
intersection lines because one cannot know which cone which 
photon comes from. 

\Figure{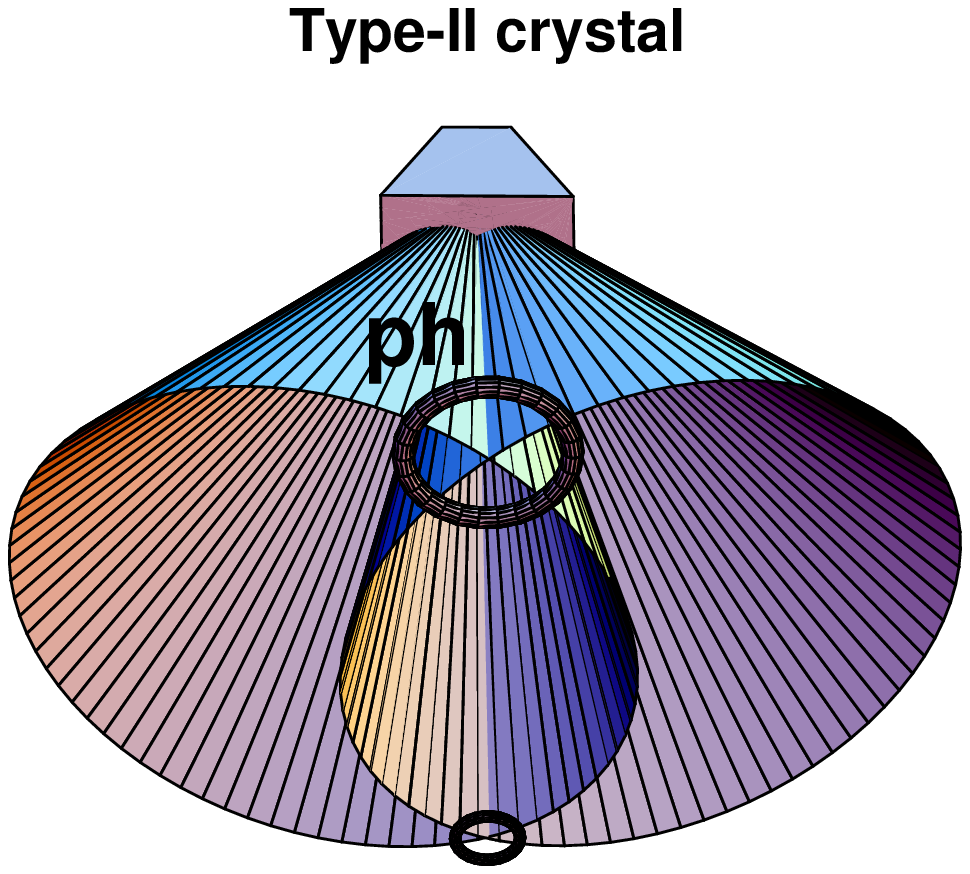}{Photons from the cones are mutually perpendicularly 
polarized and therefore the photons from the 
intersections of the cones are in a singlet--like state. 
We chose pinhole \it ph\/ \rm five times bigger than the other one 
(determined by detectors D1' and D2') so that for each photon which 
passed through the latter pinhole, its \it companion\/ \rm photon 
will pass through \it ph\/\rm.}

\lhead[]{\fontsize{9pt}{15pt}\selectfont\bf
198\hfill Mladen Pavi\v ci\'c}
\lfoot{}
\cfoot{}
\rfoot{}
\setlength{\headrulewidth}{0pt}
\setlength{\footrulewidth}{0pt}

\medskip

By an appropriate preparation one can 
entangle them in a singlet--like state.~[15] Their state 
is therefore 
\begin{eqnarray}
|\Psi\rangle={1\over\sqrt2}\bigl(|1_x\rangle_1|1_y\rangle_{1'}\>-
\>\,|1_y\rangle_1|1_x\rangle_{1'}\bigr)
\otimes{1\over\sqrt2}\bigl(|1_x\rangle_2|1_y\rangle_{2'}\>-
\>\,|1_y\rangle_2|1_x\rangle_{2'}\bigr)\,.\label{eq:4-state}
\end{eqnarray}
The outgoing electric--field operators describing photons which 
pass through beam splitter BS and through polarizers P1' and P2' 
(oriented at angles $\theta_{1'}$ and $\theta_{2'}$, respectively) 
and are detected by detectors D1' and D2' will thus read 
(see Fig.~3)
\begin{eqnarray}
\hat E_{1'}&=&\left(\hat a_{1'x}t_x\cos\theta_{1'}+\hat
a_{1'y}t_y\sin\theta_{1'}\right)
e^{i{\mathbfit k}_{1'}\cdot
{\mathbfit r}_{1'}-
i\omega_{1'}(t-t_{1'}-\tau_{1'})}\nonumber\\
&&+\ i\left(\hat a_{2'x}r_x\cos\theta_{1'}+\hat
a_{2'y}r_y\sin\theta_{1'}\right)
e^{i\tilde{\mathbfit k}_{2'}\cdot
{\mathbfit r}_{1'}-
i\omega_{2'}(t-t_{2'}-\tau_{1'})}\,,\label{eq:D2'}
\end{eqnarray}
where $t_x^2$, $t_y^2$ are transmittances, $r_x^2$, $r_y^2$ are 
reflectances, $t_j$ is time delay after which photon $j$ reaches 
BS, $\tau_{1'}$ is time delay between BS and D1',
and $\omega_{j}$ is the frequency of photon $j$. 
The annihilation operators act as follows: 
${\hat a}_{1x}|1_{x}\rangle_{1'}=|0_{x}\rangle_{1'}$, \ 
${\hat a}_{1x}|0_{x}\rangle_{1'}=0$. $E_{2'}$ is defined 
analogously. Operators describing photons which pass 
through polarizers P1 and P2 (oriented at angles $\theta_1$ 
and $\theta_2$, respectively), and through Pockels cells and 
are detected by detectors D1 and D2 will thus read 
\begin{eqnarray}
\hat E_1=
(\hat a_{1x}\cos\theta_1+\hat a_{1y}\sin\theta_1)
e^{-i\omega_1t_1}\,.\label{eq:D1}
\end{eqnarray}
$E_2$ is defined analogously. 

\lhead[]{\fontsize{9pt}{15pt}\selectfont\bf
Event--Ready Entanglement Preparation\hfill 199}
\chead{}
\rhead[]{}
\lfoot{}
\cfoot{}
\rfoot{}
\setlength{\headrulewidth}{0pt}
\setlength{\footrulewidth}{0pt}

\Figure{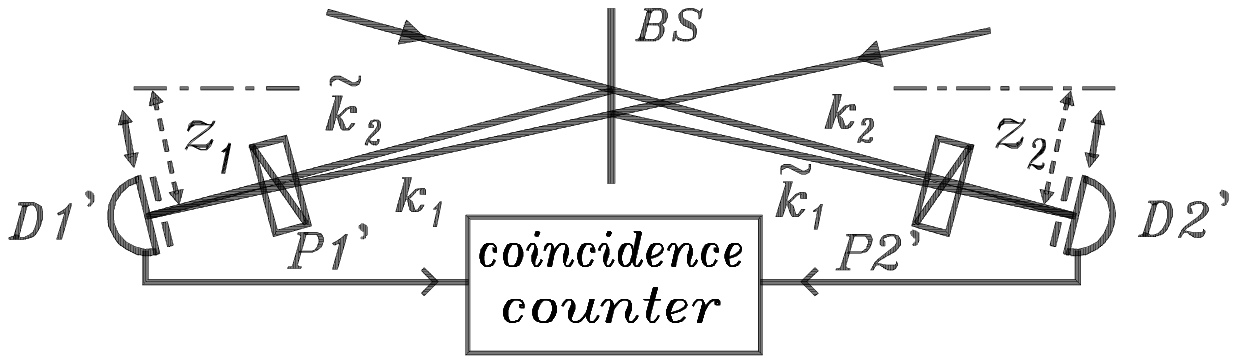}{Beam splitter BS detail from Fig.~1.}

\medskip

The probability of detecting all four photons by detectors D1,
D2, D1', and D2' is thus
\begin{eqnarray}
P(\theta_{1'},\theta_{2'},\theta_1,\theta_2)
&=&\eta^2\langle\Psi|\hat E_{2'}^\dagger\hat E_{1'}^\dagger\hat E_2^
\dagger\hat E_1^\dagger
\hat E_1^{\phantom\dagger}\hat E_2^{\phantom\dagger}
\hat E_{1'}^{\phantom\dagger}\hat E_{2'}^{\phantom\dagger}|
\Psi\rangle\nonumber\\
&=&{\eta^2\over4}(A^2+B^2-2AB\cos\phi)\,,\label{eq:prob-4} 
\end{eqnarray} 
where $\eta$ is detection efficiency; 
$A=Q(t)_{11'}Q(t)_{22'}$ 
and $B=Q(r)_{12'}Q(r)_{21'}$; here 
$Q(q)_{ij}=q_x\sin\theta_i\cos\theta_j-q_y\cos\theta_i
\sin\theta_j$; $\phi=(\tilde{\mbox{\boldmath$k$}}_2-
\mbox{\boldmath$k$}_1)\cdot\mbox{\boldmath$r$}_1+
(\tilde{\mbox{\boldmath$k$}}_1-
\mbox{\boldmath$k$}_2)\cdot\mbox{\boldmath$r$}_2=2\pi(z_2-z_1)/L$; 
here $L$ is the spacing of the interference fringes 
(see Fig.~3). $\phi$ can be changed by moving the detectors 
transversally to the incident beams. 
Data for this expression are collected by 
detectors D1' and D2' whose openings are not points but have 
a certain width $\Delta z$. Therefore, in order to obtain a  
realistic probability we 
integrate Eq.~(\ref{eq:prob-4}) over $z_1$ and $z_2$ over 
$\Delta z$ to obtain 
\begin{eqnarray}
P(\theta_{1'},\theta_{2'},\theta_1,\theta_2)
&=&{\eta^2\over4}\int\limits_{z_1-{\Delta z\over 2}}
\limits^{z_1+{\Delta z\over 2}}
\int\limits_{z_2-{\Delta z\over2}}\limits^{z_2+{\Delta z\over 2}}
\Bigl[A^2+B^2-2AB\cos\bigl[{2\pi(z_2-z_1)\over L}\bigr]\Bigr]
dz_1dz_2\nonumber\\
&=&{\eta^2\over4}(A^2+B^2-v2AB\cos\phi)\,,\label{eq:nu}
\end{eqnarray} 
where $v=\bigl[\sin(\pi\Delta z/L)/(\pi\Delta z/L)\bigr]^2$ 
is the \it visibility\/ \rm of the coincidence counting. 
We assume the near normal incidence at BS so as to have 
$r_x^2=r_y^2=R$ and $t_x^2=t_y^2=T=1-R$. Next we assume a 
symmetric position of detectors D1' and D2' with respect to 
BS and the photons paths from the middle of the crystals so as to 
obtain $\phi=0$.~[17,21] Representing photons by a 
Gaussian amplitude distribution of energies we have shown in 
Ref.~[21] that the visibility is reduced when the condition 
$\omega_{1'}=\omega_{2'}$ is not perfectly matched and when the 
coincidence detection time is not much smaller than the coherence time.  
We meet the latter demand by using a subpicosecond laser pump beam and 
the former by reducing the size of the detector (D1' and D2') pinholes. 
By reducing the size of the detector pinholes we reduce the number of 
events detected by D1 and D2 but, on the other hand, this enables 
us to increase visibility of the Bell pairs at D1 and D2 by sizing  
pinholes $ph$ (see Fig.~2) so as to make solid angles five 
times wider than the pinholes of D1' and D2'. (Cf.~Joobeur, Saleh, and 
Teich.~[13]) Alternatively, we can put $\omega_0/2$ filters 
($\omega_0$ is the frequency of the pumping beam) in front of    
detectors D1 and D2 and drop the pinholes $ph$ altogether.

\lhead[]{\fontsize{9pt}{15pt}\selectfont\bf
200\hfill Mladen Pavi\v ci\'c}
\lfoot{}
\cfoot{}
\rfoot{}
\setlength{\headrulewidth}{0pt}
\setlength{\footrulewidth}{0pt}

Let us now see in which way and when are all photons entangled. 
For $R=T=1/2$ and $v=1$ the probability Eq.~(\ref{eq:nu}) reads as 
\begin{eqnarray}
P(\theta_{1'},\theta_{2'},\theta_1,\theta_2)
={1\over4}(A-B)^2=
{1\over16}\sin^2(\theta_{1'}-\theta_{2'})\sin^2(\theta_1-\theta_2)\,.
\label{eq:l-r} 
\end{eqnarray}
and if we take away polarizers P1' and P2' the following maximal 
entanglement survives: $P(\infty',\infty',\theta_1,\theta_2)=
{1\over8}\sin^2(\theta_1-\theta_2)$. For an asymmetrical BS, however, 
if we take away polarizers P1' and P2', we obtain only partially 
entangled state 
\begin{eqnarray}
P(\infty',\infty',\theta_1,\theta_2)=
{1\over4}[(T-R)^2+2TR\sin^2(\theta_1-\theta_2)]\,.
\label{eq:par-ent} 
\end{eqnarray} 

Thus, in order to obtain (non)maximal entangled state for an 
asymmetrical beam splitter it is necessary to orient polarizers P1' 
and P2' so as to obtain a corresponding ``entangled'' 
probability given by Eq.~(\ref{eq:prob-4}). For example, for
$\phi=0^\circ$, $\theta_{1'}=90^\circ$, and $\theta_{2'}=0^\circ$,  
Eq.~(\ref{eq:prob-4}) projects out the following (non)maximal 
singlet--like probability:  
\begin{eqnarray}
P(\theta_{1},\theta_{2})&=&\eta^2s
(\cos^2\theta_{1}\sin^2\theta_{2} - 
2v\rho\cos\theta_{1}\sin{\theta_{1}}\cos\theta_{2}\sin{\theta_{2}} + 
\rho^2\cos^2\theta_{2}\sin^2\theta_{1})\nonumber\\
&\equiv &\eta^2p(\theta_{1},\theta_{2})\label{eq:r-prob}
\end{eqnarray}
where $s=T^2/(R^2+T^2)$, $\rho=R/T$, and where we multiplied 
Eq.~(\ref{eq:nu}) by 4 for other three possible coincidence 
detections [($\theta_{1'}^{\phantom{\perp}}$,$\theta_{2'}^\perp$), 
($\theta_{1'}^\perp$,$\theta_{2'}^{\phantom{\perp}}$), and 
($\theta_{1'}^\perp$,$\theta_{2'}^\perp$)] at BS and by 
$(R^2+T^2)^{-1}$ for photons emerging from the same side of BS. 

The singles--probability of detecting a photon by D1  is
\begin{eqnarray}
P(\theta_1)= 
\eta s(\cos^2\theta_1+\rho^2\sin^2\theta_1)\equiv \eta p(\theta_1)\,
.\label{eq:one} 
\end{eqnarray}
Analogously, the singles--probability of detecting a photon by D2 is 
\begin{eqnarray}
P(\theta_2)= 
\eta s(\sin^2\theta_2+
\rho^2\cos^2\theta_2)\equiv \eta p(\theta_2)\,.\hfill\label{eq:one'} 
\end{eqnarray}

\lhead[]{\fontsize{9pt}{15pt}\selectfont\bf
Event--Ready Entanglement Preparation\hfill 201}
\chead{}
\rhead[]{}
\lfoot{}
\cfoot{}
\rfoot{}
\setlength{\headrulewidth}{0pt}
\setlength{\footrulewidth}{0pt}

Introducing the above obtained probabilities into the Clauser--Horne 
inequality (\ref{eq:enh}) we obtain the following minimal efficiency 
for its violation. 
\begin{eqnarray}   
\eta={p(\theta_1')-p(\theta_2)\over 
p(\theta_1,\theta_2)-p(\theta_1,\theta_2')
+p(\theta_1',\theta_2')+p(\theta_1',\theta_2)}
\,.\hfill\label{eq:eta} 
\end{eqnarray}
This efficiency is a function of visibility $v$ and by looking at 
Eqs.~(\ref{eq:r-prob}), (\ref{eq:one}), and (\ref{eq:one'})  
we see that for each particular $v$ a different set of angles 
should minimize it. 

\medskip 

A computer optimization of angles---presented in 
Fig.~4---shows that the lower the reflectivity is, the lower is the 
minimal detection efficiency. Also, we see a  rather 
unexpected property that a low visibility does not have a significant 
impact on the violation of the Bell inequality. For example, with 
70\%\ visibility and 0.2 reflectivity of the beam splitter we 
obtain a violation with a lower detection efficiency than with  
100\%\ visibility and 0.5 ($\rho=1$) reflectivity. 

A similar calculation can be carried out for the Hardy equalities 
given at the and of Sec.~2. It can be shown that the lowest possible 
$R$, with only 5--10 standard deviations, should be taken and not the 
one which gives the greatest $P(\theta_1,\theta_2)>0$, again because 
the impact of a low visibility is the lowest when the beam splitter 
is the most asymmetrical. Thus our preselection scheme can be used 
for a loophole--free ``Hardy experiment'' as well.

\vfill\eject 

\Figuresized{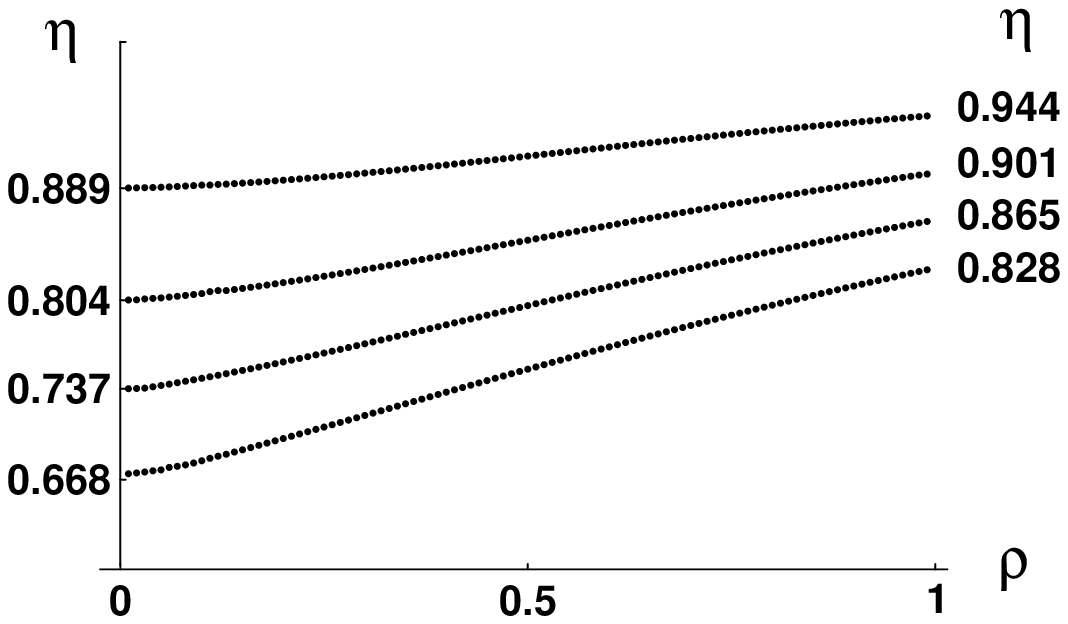}{Minimal detection efficiencies 
$\eta$ necessary for a violation of the Bell--Clauser--Horne inequality 
as functions of visibility $v$ and of $\rho=R/(1-R)$, where $R$ is the 
reflectivity of the beam splitter.}{5.4}

\lhead[]{\fontsize{9pt}{15pt}\selectfont\bf
202\hfill Mladen Pavi\v ci\'c}
\lfoot{}
\cfoot{}
\rfoot{}
\setlength{\headrulewidth}{0pt}
\setlength{\footrulewidth}{0pt}

\vfill\eject

\section{\fontsize{12pt}{12pt}\selectfont\bf Conclusion}
\label{sec:concl}

Our elaboration shows that the recently found four--photon 
entanglement~[18,21] can be used for a realization of 
loophole--free \it Bell experiments\/\rm. We propose a set--up 
which uses two simultaneous type--II downconversions into two 
singlet--like photon pairs. By combining two photons, one from 
each such singlet--like pair, at an asymmetrical beam splitter 
and detecting them in coincidence we preselect the other two 
completely independent photons into another singlet--like state---let 
us call them `\it Bell pair\/\rm'. (See Figs.~1 and 3.) Our 
calculations show that no time or space windows are imposed on 
the \it Bell pairs\/ \rm by the preselection 
procedure and this means that we can collect the photons within 
an optimal solid angle. If we take their solid angles five times 
wider than the angles of \it preselector\/ \rm photons 
(determined by the openings of detectors D1' and D2'---see Fig.~1), 
then we can collect all \it Bell pairs\/ \rm and at the same time 
keep a probability of the ``third party'' counts negligible.

\lhead[]{\fontsize{9pt}{15pt}\selectfont\bf
Event--Ready Entanglement Preparation\hfill 203}
\chead{}
\rhead[]{}
\lfoot{}
\cfoot{}
\rfoot{}
\setlength{\headrulewidth}{0pt}
\setlength{\footrulewidth}{0pt}

For our set--up we can use the result presented in 
Fig.~4 which enables a conclusive violation of Bell's inequalities 
with a detection efficiency lower than 80\%\ even when the 
visibility is under 70\%\ at the same time. If we, however, agree 
that it is physically plausible to take into account only those 
\it Bell pairs\/ \rm which are preselected by actually recorded 
detections at the beam splitter (firing of D1' and D2'), 
then we can eliminate the low visibility impact altogether.
In this case, we can set $v=1$ and for a \it different\/ \rm set of 
angles obtain a conclusive violation of Bell's inequalities and 
Hardy's equalities with still lower (under 70\%) detection 
efficiency. 

In the end, we stress that the whole device can also be used for 
delivering ready--made Bell pairs in quantum cryptography and quantum 
computation and communication.

\vbox to 0.5cm{\vfill}

\parindent=0pt
{\fontsize{12pt}{12pt}\selectfont\bf Acknowledgments} 

\medskip 

I acknowledge supports of the Alexander von Humboldt Foundation, 
and the Ministry of Science of Croatia. 

\bigskip
\bigskip
\bigskip
\bigskip

{\fontsize{12pt}{12pt}\selectfont\bf References}

\bigskip

\everypar={\parindent=0pt\hangindent=20pt\hangafter=1}

\ $\,$[1] Ardehali, M., \it Phys.~Rev.\/~\rm A {\bf 49}, 
R3143 (1994).

\ $\,$[2] Ardehali, M., \it Phys.~Rev.\/~\rm A {\bf 49}, 
R3143 (1994).

\ $\,$[3] Clauser, J.~F.~and M.~A.~Horne, \it 
Phys.~Rev.\/~\rm D {\bf 10}, 526 (1974).

\ $\,$[4] Clauser, J.~F.~and A.~Shimony,
                    \it Rep.~Prog.~Phys.\/~\rm {\bf 41}, 1881 (1978).

\ $\,$[5] Eberhard, P.~H., \it Phys.~Rev.\/~\rm A,\ 
{\bf 47}, R747 (1993).

\ $\,$[6] Fry, E.~S., T.~Walther, and S.~Li, 
                    \it Phys.~Rev.\/~\rm A {\bf 52}, 4381 (1995).

\ $\,$[7] Garg, A.~and N.~D.~Mermin,
                    \it Phys.~Rev.\/~\rm D {\bf 35}, 3831 (1987).

\ $\,$[8] Garuccio, A., \it Ann.~N.~Y.~Acad.~Sci.\/~\rm 
{\bf 755} 632 (1995).

\ $\,$[9] Hardy, L., \it Phys.~Rev.~Lett.\/~\rm {\bf 71}, 
1665 (1993). 

[10] Home, D.~and F.~Selleri, 
               \it Riv.~Nuovo Cim.\/~\rm {\bf 14}, \sl No.~9\/\rm, (1991). 

[11] Huelga, S.~F., M.~Ferrero, and E.~Santos, 
                    \it Phys.~Rev.\/~\rm A {\bf 51}, 5008 (1995).

[12] Jones, R.~T.~and E.~G.~Adelberger, 
                    \it Phys.~Rev.~Lett.\/~\rm {\bf 72}, 267 (1994).

[13] Joobeur, A., B.~E.~A.~Saleh, and M.~C.~Teich
                    \it Phys.~Rev.\/~\rm A {\bf 50}, 3349 (1994).
 
[14] Kwiat, P.~G., P.~H.~Eberhard, A.~M.~Steinberg, and R.~Y.~Chiao, 
                    \it Phys.~Rev.\/~\rm A, {\bf 49}, 3209 (1994).

[15] Kwiat, P.~G.~, K.~Mattle, H.~Weinfurter, and A.~Zeilinger, 
                    \it Phys.~Rev.~Lett.\/ \rm~{\bf 75}, 4337 (1995). 

[16] Ou, Z.~Y.~and L.~Mandel, \it Phys.~Rev.~Lett.\/~\rm {\bf 61}, 50 (1988); 

[17] Pavi\v ci\'c, M., \it Phys.~Rev.\/~\rm A {\bf 50}, 3486 (1994).

[18] Pavi\v ci\'c, M., \it J.~Opt.~Soc.~Am.\/~\rm B, {\bf 12}, 821 (1995). 

[19] Pavi\v ci\'c, M., \it Phys.~Lett.\/~\rm A {\bf 209}, 255 (1995).

[20] Pavi\v ci\'c, M.~in \it Fourth International Conference on Squeezed 
                    States and Uncertainty Relations\/\rm, 
                    (NASA CP 3322, Washington, 1996), pp.~325. 

[21] Pavi\v ci\'c, M.~and J.~Summhammer, 
                    \it Phys.~Rev.~Lett.\/~\rm {\bf 73}, 3191 (1994). 

[22] Santos, E., \it Phys.~Rev.~Lett.\/~\rm {\bf 66}, 1388 (1991). 

[23] Santos, E., \it Phys.~Rev.~Lett.\/~\rm{\bf 68}, 2702 (1992). 

[24] Santos, E., \it Phys.~Rev.\/~\rm A {\bf 46}, 3646 (1992). 

[25] Santos, E., \it Phys.~Lett.\/~\rm A {\bf 212}, 10 (1996). 

[26] Torgerson, J., D.~Branning, and L.~Mandel, 
                    \it App.~Phys.\/~\rm {\bf 60}, 267 (1995).

[27] Torgerson, J., D.~Branning, C.H.~Monken, and L.~Mandel, 
                    \it Phys.~Lett.\/~\rm A {\bf 204}, 323 (1995).

\lhead[]{\fontsize{9pt}{15pt}\selectfont\bf
204\hfill Mladen Pavi\v ci\'c}
\lfoot{}
\cfoot{}
\rfoot{}
\setlength{\headrulewidth}{0pt}
\setlength{\footrulewidth}{0pt}

\end{document}